\documentclass[letterpaper]{JHEP3}
\usepackage{bm}
\usepackage{amsmath}
\usepackage{yfonts}
\usepackage{epsfig,latexsym}
\usepackage{cite}

\newcommand{\beq}{\begin{equation}}
\newcommand{\eeq}{\end{equation}}
\newcommand{\bqa}{\begin{eqnarray}}
\newcommand{\eqa}{\end{eqnarray}}

\title{On the collision of two shock waves in $AdS_5$}

\preprint{INT PUB 08-06\\
MIT-CTP 3939\\
YITP-08-18}

\author{Daniel Grumiller \\
Center for Theoretical Physics, Massachusetts Institute of Technology, \\
77 Massachusetts Ave., Cambridge, MA, 02139, USA \\
E-mail: \email{grumil@lns.mit.edu}}
\author{Paul Romatschke\\
Institute for Nuclear Theory, University of Washington, \\
Box 351550, Seattle, WA, 98195, USA\\
and\\
Yukawa Institute for Theoretical Physics, Kitashirakawa-Oiwakecho,\\
Sakyo, Kyoto 606-8502, JAPAN\\
E-mail: \email{paulrom@phys.washington.edu}
}

\date{\today}

\abstract{We consider two ultrarelativistic shock waves propagating and
colliding in five-dimensional Anti-de-Sitter spacetime.
By transforming to Rosen coordinates, we are able to find the
form of the metric shortly after the collision. 
Using holographic renormalization, we calculate 
the energy-momentum tensor on the boundary of 
AdS space for early times after the collision.
Via the gauge-gravity duality, this gives some insights on
bulk dynamics of systems created by high energy 
scattering in strongly coupled gauge theories.
We find that Bjorken boost-invariance is explicitely
violated at early times and we obtain an estimate for the thermalization
time in this simple system.}

\begin{document}

\section{Introduction}

With the advent of high energy colliders with collision energies in the TeV range, 
progress in understanding the problem of ultrahigh energy 
particle scattering now involves knowledge about the dynamics of
Quantum-Chromodynamics (QCD) above or close to the deconfinement scale.
For instance, a new state of matter of deconfined quarks and gluons 
with fluid-like properties seems to have been created in Au+Au collisions 
at total collision energies of $\sim 40$ TeV
at the Relativistic Heavy-Ion Collider (RHIC)
\cite{Adcox:2004mh,Back:2004je,Arsene:2004fa,Adams:2005dq}. 
The upcoming Large Hadron
Collider (LHC) will collide protons at $\sim 14$ TeV and lead nuclei at up to 
$\sim 1100$ TeV. Although not a collider, the Pierre-Auger 
observatory probes collision energies much beyond that, at 
up to $\sim 10^{8}$ TeV.

QCD is a complicated theory to solve, especially when
asking about real-time dynamics at energies close to the deconfinement
scale. Traditionally one could only resort to weak coupling 
approaches, which due to asymptotic freedom
are guaranteed to work well for bulk systems 
having extremely high energy densities
(though not necessarily for those probed by RHIC and the LHC).
In particular, 
phenomenological descriptions of RHIC data by applying hydrodynamic
simulations with low viscosity \cite{Romatschke:2007mq}
or none at all \cite{Teaney:2000cw,Huovinen:2001cy,Hirano:2002ds}
seem to suggest that at these energy densities,
the system is not weakly coupled \cite{Shuryak:2004cy,Lee:2005gw}.

Recently, the conjectured duality between strongly coupled gauge theories
and gravity \cite{Maldacena:1997re} has opened up a new window 
for studying strong coupling phenomena
in a range of different gauge theories (although a dual description 
of QCD remains elusive to date). Concerning real-time dynamics,
a lot of progress has been made in calculating transport coefficients
for hydrodynamics in strong coupling, such as shear viscosity
\cite{Policastro:2001yc}.
However, these calculations probe the response of a static medium
at finite temperature, which -- while important for near-equilibrium
dynamics -- do not give insights into the early, far from equilibrium
stages of a high energy particle collision.

From the gauge theory point of view this earliest stage following
the collision has to describe the transition of the system to
an equilibrium state with a well-defined temperature
(if the system lives long enough), which is referred to as thermalization.
The gravity dual picture of thermalization is 
the formation of a black hole in the bulk, its Hawking temperature 
being identified with the temperature in the gauge theory on the boundary
of the Anti-de-Sitter (AdS) space. Thus the problem of thermalization
in strongly coupled gauge theories becomes related to the problem of
black hole formation, which has been noted before 
\cite{Nastase:2005rp} (see also \cite{Castorina:2007eb,Castorina:2008gf}
for a different proposal on thermalization involving black holes).
%

As an aside, it should be pointed out that thermalization in a high 
energy particle collision is not guaranteed, and even if once 
achieved, not easy to maintain due to the rapid (initially 
one-dimensional) expansion of the system. 

In an inspiring work Janik and Peschanski \cite{Janik:2005zt}
showed that an expanding thermal system corresponded to a gravity
dual where the black hole was moving in the fifth dimension.
Subsequent work on gauge-gravity duality in expanding systems
continues to clarify the near-equilibrium late-time behavior
of high energy particle collision duals 
\cite{Janik:2006gp,Nakamura:2006ih,Sin:2006pv,Janik:2006ft,Lin:2006rf,Friess:2006kw,Kajantie:2006ya,Heller:2007qt,Kajantie:2007bn,Kim:2007ut,Benincasa:2007tp}.

Little is known about the dynamics in strongly 
coupled gauge theories shortly after a high energy collision, e.g.~far from equilibrium. Notable exceptions are studies assuming
independence from longitudinal dynamics \cite{Kovchegov:2007pq},
treating one space dimension instead of three \cite{Kajantie:2008rx}
and 
a characterization of the dual horizon structure following 
a collision of two already deconfined plasmas \cite{Amsel:2007cw}. 
We differ from \cite{Amsel:2007cw} by using a simpler model
for the incident states, which allows us to calculate the
energy-momentum tensor analytically. Differences of our 
approach to Refs.~\cite{Kovchegov:2007pq,Kajantie:2008rx} will
be discussed below.

In this article we shall consider the problem of two colliding
infinite sheets of matter in ${\mathcal N}=4$ SYM in the large $N_c$ and
large 't~Hooft coupling limit. Via the gauge-gravity duality
we reinterpret the problem as the collision of two shock waves
on the boundary of a five-dimensional AdS space. Solving for
the five dimensional metric and using holographic
renormalization, one thus is able to extract information about
the real time dynamics of the energy momentum tensor of the gauge
theory after the collision. 

Our work is organized as follows: in Section \ref{sec:one} we construct the line-element for two colliding shock-waves in Rosen coordinates. Section \ref{sec:two} deals with holographic renormalization: we construct there the energy momentum tensor perturbatively in proper time and check that it is covariantly conserved and traceless. We provide a physical interpretation of our results in Section \ref{sec:three}, and put them into the perspective of the literature.

\section{Setup and Solution}
\label{sec:one}

Following \cite{Janik:2005zt}, we consider for ${\mathcal N}=$ SYM in
the strong coupling, large $N_c$ regime, an energy-momentum
tensor (EMT) of the single-shock-wave-form
\beq
T_{++}=0,\quad T_{+-}=0\,, \quad T_{--}=\mu_1 \,\delta(x^-)\,,\quad T_{xx}=T_{yy}=0
\label{EMTpre1}
\eeq
in light-cone coordinates $x^\pm=\frac{t\pm z}{\sqrt{2}}$.
This form serves as a toy model for a large  
particle moving nearly at the speed of light along the $x^+$ direction
with transverse energy density $\frac{dE}{d{\bf x}_\perp}=\mu_1$. The attribute ``large'' refers to the perpendicular directions, ${\bf x_\perp}=(x,y)$.
In the gravity dual description, in Fefferman-Graham coordinates
\cite{Fefferman}
this corresponds to the $AdS_5$ line element with a single shock-wave 
\beq
ds^2=\frac{-2 dx^+ dx^-+ \mu_1 z^4 \delta(x^-) dx^{-2}+d{\bf x_\perp}^2
+dz^2}{z^2}\,,
\label{EMTpre2}
\eeq
which is an exact solution of the Einstein equations with negative cosmological constant (normalized here to $\Lambda=-6$)
\beq
R_{\mu \nu}-\frac{1}{2}g_{\mu \nu} R-6 g_{\mu \nu}=0\,.
\label{eq:Eeq}
\eeq

In the following we shall consider the collision of two
shock-waves of the form (\ref{EMTpre1},\ref{EMTpre2}),
corresponding to an EMT before the collision 
($t<0$) of
\beq
T_{++}=\mu_2 \,\delta(x^+)\,,
\quad T_{+-}=0\,, 
\quad T_{--}=\mu_1 \,\delta(x^-),\quad T_{xx}=T_{yy}=0\,.
\label{EMTpre3}
\eeq
This problem was posed originally in Ref.\cite{Janik:2005zt}.
The simple question we want to address is: what is the form of the
EMT after the collision, and in particular
in the forward light-cone ($x^\pm>0$)?

This setup of particle collisions in strongly coupled
gauge theories mirrors closely that of Kovner, McLerran and Weigert
\cite{Kovner:1995ja} who treated collisions using
classical Yang-Mills dynamics. Starting from a charge current
$J^\mu=\delta^\mu_{\mp} \delta(x^\pm) \rho_{1,2}({\bf x}_\perp)$ and solving
the classical Yang-Mills equations $D_\mu F^{\mu \nu}=J^\nu$
they showed that the resulting gauge fields (and hence the EMT)
were functions of the product $x^+ x^-$ only, and therefore
``boost-invariant'' in the sense of Bjorken \cite{Bjorken:1982qr}.
Quantum fluctuations are expected to break this boost-invariance
\cite{Fukushima:2006ax}, since 
even tiny fluctuations are unstable to exponential
growth \cite{Romatschke:2005pm}. Our model (\ref{EMTpre3})  
in some sense 
corresponds to the simple case 
$\rho_{1,2}({\bf x}_\perp)=\mu_{1,2}={\rm const.}$, but is
extendable to a situation where the $\rho$'s are taken from
the Color Glass Condensate framework 
\cite{McLerran:1993ni,McLerran:1993ka}. We shall report
on this interesting generalization in a subsequent work \cite{inprep}.

In order to calculate the energy momentum tensor in the forward 
light-cone for strongly coupled ${\mathcal N}=4$ SYM, let
us first use the coordinate transformation\footnote{In some of the expressions below positive powers of the $\theta$-function will appear. For positive and negative argument the square of the $\theta$-function is equivalent to the $\theta$-function, so the only complication arises if the argument vanishes. However, it turns out that all expressions of the type $\theta^n(x)$ are multiplied by $x^m$ with positive $m$, and thus this complication and the associated ambiguity of defining $\theta^n(0)$ is of no relevance here.} \cite{painanddeath}
\beq
x^+=u+\frac{1}{2} \mu_1 \theta(v)\ \tilde{z}^4+2 \mu_1^2\ v\ \theta^2(v)\
\tilde{z}^6\,,\quad
x^-=v\,,\quad
z=\tilde{z}+2 \mu_1\ v\ \theta(v) \tilde{z}^3\,,
\label{firstcotra}
\eeq
where $\theta(v)$ is the Heaviside step function,
to bring Eq.~(\ref{EMTpre2}) into the so-called Rosen form,
\beq
ds^2=\frac{-2 du dv +d{\bf x_\perp}^2+
\left[1+6 \mu_1 \tilde{z}^2\ v\ \theta(v)\right]^2 d\tilde{z}^2}{\tilde{z}^2
\left[1+2 \mu_1 \tilde{z}^2\ v\ \theta(v)\right]^2}\,.
\label{sshockrosen}
\eeq
This form is advantageous since it implies a metric that is 
continuous 
across the light-like hypersurface $v=0$.
Indeed, since one can do a similar transformation for the
second shock wave, the precollision line element can be written
as a simple superposition of the two (c.f. \cite{painanddeath}),
\beq
ds^2=\frac{-2 du dv +d{\bf x_\perp}^2+
\left[1+6 \mu_1 \tilde{z}^2\ v\ \theta(v)+6 \mu_2 \tilde{z}^2\ u\ \theta(u)
\right]^2 d\tilde{z}^2}{\tilde{z}^2
\left[1+2 \mu_1 \tilde{z}^2\ v\ \theta(v)+2 \mu_2 \tilde{z}^2\ u\ \theta(u)
\right]^2}\,.
\label{superpo}
\eeq

Since the metric has to be continuous and piece-wise differentiable 
in these coordinates, all corrections to the above line element after collision have to be
proportional to $uv\, \theta(u)\, \theta(v)$ (see appendix
\ref{mappendix} for a more detailed discussion). 
The calculation of these corrections is most conveniently done
by introducing the coordinates of proper time $\tilde{\tau}=\sqrt{2 u v}$
and space-time rapidity $\tilde{\eta}=\frac{1}{2}\ln{\frac{u}{v}}$.
In these coordinates the hyper-surface spanned by $u=v=0$ becomes
$\tilde\tau=0$ and the condition $\theta(u) \theta(v)\neq 0$ translates
into $\tilde\tau$ being real and positive. Introducing $\mu=\sqrt{2 \mu_1 \mu_2}$ 
and $Y=\frac{1}{2} \ln{\frac{\mu_1}{\mu_2}}$,
we then make the following ansatz for the
line element after the collision
\bqa
ds^2&=&\frac{-\left[1+K(\tilde\tau,\tilde\eta,\tilde{z})\right]d\tilde\tau^2
+\left[1+L(\tilde\tau,\tilde\eta,\tilde{z})\right]\tilde\tau^2 d\tilde\eta^2 
+\left[1+H(\tilde\tau,\tilde\eta,\tilde{z})\right]d{\bf x_\perp}^2}{\tilde{z}^2
\left[1
+2 \tilde{z}^2 \mu \tilde\tau \cosh(Y-\tilde\eta)
\right]^2}\nonumber\\
&&\hspace*{4cm}
+\frac{\left[1+M(\tilde\tau,\tilde\eta,\tilde{z})\right]
\left[1+6 \tilde{z}^2 \mu \tilde\tau \cosh(Y-\tilde\eta)
\right]^2 d\tilde{z}^2}{\tilde{z}^2
\left[1+2 \tilde{z}^2 \mu \tilde\tau\cosh(Y-\tilde\eta)\right]^2}\,,
\label{oursol}
\eqa
where $K,L,H,M$ are functions that vanish at $\tilde\tau=0$ and have
to be determined by solving the Einstein equations \eqref{eq:Eeq}.

Determining $K,L,H,M$ for all values of $\tilde\tau$ may be possible 
with existing numerical methods 
\cite{Pretorius:2005gq,Husa:2007zz,inprep}. Finding
full analytical solutions is much harder, so we limit
ourselves to the restricted regime of early times $\tilde\tau\ll 1$.
For this regime we use a power series ansatz in $\tilde\tau$
for the functions $K,L,H,M$ and determine the coefficients 
by solving the Einstein equations order by order
in proper time. This is readily done with {\tt GRTensor}
\cite{GRTensor}.
One finds
\bqa
K(\tilde\tau,\tilde\eta,\tilde{z})&=&c_1 \mu^2 \tilde\tau^2 \tilde{z}^4
+\frac{182+10 c_1}{3}\mu^3 \tilde\tau^3 \tilde{z}^6 \cosh[Y-\tilde\eta]
-\frac{5+c_1}{3}\mu^2 \tilde\tau^4 \tilde{z}^2
\nonumber\\
&&+\frac{838+160 c_1+c_1^2+45 c_2}{9} \mu^4 \tilde\tau^4 \tilde{z}^8
-\frac{154+2c_1}{3} \mu^4 \tilde\tau^4 \tilde{z}^8 \cosh[2(Y-\tilde\eta)]
+{\mathcal O}(\tilde\tau^5)\nonumber\\
L(\tilde\tau,\tilde\eta,\tilde{z})&=&\frac{-16+c_1}{3} \mu^2 \tilde\tau^2 \tilde{z}^4
+\frac{94+2 c_1}{3} \mu^3 \tilde\tau^3 \tilde{z}^6 \cosh[Y-\tilde\eta]
-\frac{5+c_1}{3}\mu^2 \tilde\tau^4 \tilde{z}^2\nonumber\\
&&+c_2 \mu^4 \tilde\tau^4 \tilde{z}^8
-\frac{806+10 c_1}{15} \mu^4 \tilde\tau^4 \tilde{z}^8 \cosh[2(Y-\tilde\eta)]
+{\mathcal O}(\tilde\tau^5)\nonumber\\
H(\tilde\tau,\tilde\eta,\tilde{z})&=&-2 \mu^2 \tilde\tau^2 \tilde{z}^4
-\frac{22+2 c_1}{3}\mu^3 \tilde\tau^3 \tilde{z}^6 \cosh[Y-\tilde\eta]
-\frac{5+c_1}{3}\mu^2 \tilde\tau^4 \tilde{z}^2\nonumber\\
&&-\frac{16+2 c_1}{3} \mu^4 \tilde\tau^4 \tilde{z}^8
+\frac{8-2 c_1}{3} \mu^4 \tilde\tau^4 \tilde{z}^8 \cosh[2(Y-\tilde\eta)]
+{\mathcal O}(\tilde\tau^5)\nonumber\\
M(\tilde\tau,\tilde\eta,\tilde{z})&=&16 \mu^2 \tilde\tau^2 \tilde{z}^4
-\frac{244-4 c_1}{3} \mu^3 \tilde\tau^3 \tilde{z}^6 \cosh[Y-\tilde\eta]
+\frac{10+2 c_1}{3}\mu^2 \tilde\tau^4 \tilde{z}^2\nonumber\\
&&+\frac{1076+4 c_1}{3} \mu^4 \tilde\tau^4 \tilde{z}^8
+\frac{824-8 c_1}{3}\mu^4 \tilde\tau^4 \tilde{z}^8 \cosh[2(Y-\tilde\eta)]
+{\mathcal O}(\tilde\tau^5)\,,
\label{solfunc}
\eqa
where $c_1,c_2$ are freely choosable integration constants. 
Amusingly, all terms of ${\mathcal O}(\tilde\tau^4 \tilde z^2)$ 
in \eqref{solfunc} cancel for the choice $c_1=-5$; 
however, at higher orders there is no similar freedom,
and e.g. the ${\mathcal O}(\tilde\tau^7 \tilde z^2)$ terms 
cannot be canceled.
%
It is straightforward, but a little tedious 
to calculate $K,L,H,M$ to arbitrary order in $\tilde\tau$. We have performed
the calculation up to (including) ${\mathcal O} (\tilde\tau^{10})$, but 
believe little insight can be gained by spelling out this solution
here. 

The constants $c_1, c_2$ reflect a residual gauge freedom, 
\beq
{\cal L}_\xi \,g_{\mu\nu} = \xi^\alpha \partial_\alpha g_{\mu\nu} + g_{\mu\alpha} \partial_\nu \xi^\alpha + g_{\alpha\nu} \partial_\mu \xi^\alpha\,.
\label{eq:lie}
\eeq
To exhibit the property of $c_1$ as parameter of residual gauge transformations we focus here on the leading order, i.e., we consider only terms up to (including) order $\tilde\tau^2$ (with the exception of $g_{\tilde\eta \tilde\eta}$, which has to be considered up to order $\tilde\tau^4$). With the generator 
\beq
\xi^{\tilde\tau} = \frac{c_1 \mu^2 \tilde\tau^3 \tilde{z}^4}{6}\,,\qquad \xi^\mu = 0 \quad{\rm otherwise}
\label{eq:rgt}
\eeq
the gauge transformation acts on the metric as follows:
\beq
{\cal L}_\xi \,g_{\tilde\tau \tilde\tau} 
= -\frac{1}{\tilde{z}^2}\,c_1\mu^2\tilde\tau^2\tilde{z}^4 
+ {\mathcal O}(\tilde\tau^3)\,, \qquad
{\cal L}_\xi \,g_{\tilde\eta \tilde\eta} = \frac{\tilde\tau^2}{\tilde{z}^2}\,\frac{c_1\mu^2\tilde\tau^2\tilde{z}^4}{3} + {\mathcal O}(\tilde\tau^5) \label{eq:bla}
\eeq
All other components of the metric are either not influenced at all, or only at order $\tilde\tau^3$. Comparison of 
\eqref{eq:bla} 
with \eqref{solfunc}  to order $\tilde\tau^2$ reveals that the terms generated by the residual gauge transformation \eqref{eq:lie} with generator \eqref{eq:rgt} are precisely the $c_1$-dependent terms in \eqref{solfunc}. This shows clearly that the freedom to choose $c_1$ corresponds to a residual gauge freedom of the (partially) gauge fixed metric \eqref{oursol}, and thus we should expect that physical quantities, like the EMT, are independent from $c_1$. We shall demonstrate this in the next Section. A similar analysis applies to higher orders in $\tilde\tau$, but the corresponding generator of residual gauge transformations is considerably more complicated than \eqref{eq:rgt}.

\section{Holographic Renormalization}
\label{sec:two}

Having determined the solution to the metric for short times
after the collision in the previous Section, we focus now
on extracting information about the gauge theory EMT in
this Section. Holographic renormalization \cite{deHaro:2000xn}
gives a simple prescription to obtain the EMT once the metric
is in the Fefferman-Graham form, 
\beq
ds^2=\frac{g_{ij} dx^i dx^j}{z^2}
+\frac{z^4 T_{ij} dx^i dx^j}{z^2}+\sum_{n=0}^{\infty}
\frac{z^{6+2 n} h_{ij}^{(n)} dx^i dx^j}{z^2}\,,
\label{holren}
\eeq
where $i$ collectively denote the coordinates on the AdS boundary ($z=0$),
$g_{ij}$ is the metric on the boundary (assumed to be Minkowski)
and $T_{ij}$ the gauge theory EMT.
We can achieve to bring Eq.~(\ref{oursol}) with (\ref{solfunc}) into the form (\ref{holren}) by 
the coordinate transformation
\bqa
\tilde{\tau}&=&\tau+\sum_{n=0}^{\infty} t_n(\tau,\eta) z^{4+2 n}\nonumber\\
\tilde{\eta}&=&\eta+\sum_{n=0}^{\infty} e_n(\tau,\eta) z^{4+2 n}\nonumber\\
\tilde{z}&=&z+\sum_{n=0}^\infty a_n(\tau,\eta) z^{3+2n}\,,
\eqa
where the $z=0$ transformation in $\tilde{\tau},\tilde\eta$ has been
chosen such that $g_{ij}$ maintains the form
$$g_{ij}dx^i dx^j=-d\tau^2+\tau^2 d\eta^2 + d{\bf x}_\perp^2\ .$$
One finds that the matching to lowest orders requires
\bqa
a_0(\tau,\eta)&=&-2 \mu \tau \cosh[Y-\eta]-\frac{5+c_1}{6}\mu^2 \tau^4
-\mu^3 \tau^7 \cosh[Y-\eta]+{\mathcal O}(\tau^{8})\nonumber\\
t_0(\tau,\eta)&=&\frac{1}{4}\partial_\tau a_0(\tau,\eta)\nonumber\\
e_0(\tau,\eta)&=&-\frac{1}{4\tau^2}\partial_\eta a_0(\tau,\eta)
\eqa
and in turn leads to the EMT
\bqa
T_{\tau \tau}&=&\mu^2 \tau^2-3 \mu^3 \tau^5 \cosh[Y-\eta]
+\frac{1}{24} \mu^4 \tau^8 \left(107+90 \cosh[2(Y-\eta)]\right)
+{\mathcal O}(\tau^{10})\nonumber\\
T_{\eta \tau}&=&-3 \mu^3 \tau^6 \sinh[Y-\eta]
+\frac{45}{4} \mu^4 \tau^9 \sinh[2 (Y-\eta)]
+{\mathcal O}(\tau^{10})\nonumber\\
\tau^{-2} T_{\eta \eta}&=&-3 \mu^2 \tau^2+21 \mu^3 \tau^5
\cosh[Y-\eta]
-\frac{3}{8}\mu^4 \tau^8 \left(107+150 \cosh[2(Y-\eta)]\right)
+{\mathcal O}(\tau^{10})\nonumber\\
T_{xx}=T_{yy}&=&2 \mu^2 \tau^2-12 \mu^3 \tau^5 \cosh[Y-\eta]
+\frac{5}{24} \mu^4 \tau^8 \left(107+144 \cosh[2(Y-\eta)]\right)
+{\mathcal O}(\tau^{10})\ .\nonumber\\
\label{EMTresult}
\eqa
Note that the gauge theory EMT is independent from
the residual gauge parameters $c_1,c_2,\ldots$.
Moreover, it obeys $T^\mu_\mu=0$ and 
$\nabla_\mu T^{\mu \nu}=0$, as it should,
where we recall that $\nabla_\mu$ is the covariant
derivative with respect to the metric $g_{ij}$.

\section{Physics Interpretation}
\label{sec:three}

Our main result, Eq.~(\ref{EMTresult}), gives the energy momentum tensor
for short times after the collision of two sheets of matter
in a strongly coupled gauge theory.
A few remarks are in order: a non-vanishing off-diagonal element
of $T_{i \tau}$ in $\tau,\eta$ coordinates means there is a flow
of energy in the direction $i$, so the EMT is not in its
local rest-frame. Locally, the EMT may always be brought into its 
rest-frame by a Lorentz boost (e.g.~$\eta\rightarrow \eta+\phi$).
However, the form of Eq.~(\ref{EMTresult}) is such that the EMT
may not be brought into its rest-frame globally (i.e.~for all
$\eta$ simultaneously), since it is explicitly dependent on $\eta$.
Put differently, the EMT (\ref{EMTresult}) is not boost-invariant
in the sense of Bjorken. This is a major difference to the 
result found when treating the gauge interaction as classical 
Yang-Mills fields (see discussion in the introduction).
It may still be possible that boost-invariance is recovered
(at least approximately) at late times, where our 
solution breaks down. The full solution obtained
in a lower dimensional model 
\cite{Kajantie:2008rx}, where boost-invariance 
(violated by construction at early times)
is restored at late times, seems to suggest
that this is the case.

Nevertheless, it is interesting to discuss the time
behavior of the EMT for one particular rest-frame, 
e.g.~``central rapidity'' $\eta=Y$. Then the individual diagonal
components of the EMT have the interpretation of (local)
energy density ($T_{\tau \tau}$), effective longitudinal pressure
($\tau^{-2} T_{\eta \eta}$) and effective transverse pressure
($T_{xx}=T_{yy}$). Keeping in mind that the result (\ref{EMTresult}) 
is valid only for $\tau>0$ since it does not contain
the original discontinuities (\ref{EMTpre3}), it is interesting
to note that all components of $T_{\mu \nu}$ are very small
initially and grow only proportional to $\tau^2$ (although
with negative sign in the case of the longitudinal pressure). 
We believe that this may be due to the simplicity of our model, and
more specifically to the absence of transverse (${\bf x}_\perp$) dynamics in
our ansatz for $\mu$ in (\ref{EMTpre3}). Along the same
lines, Ref.~\cite{Kajantie:2008rx} found that
$T_{\mu \nu}$ would vanish for all times had one started
with (\ref{EMTpre3}) and completely eliminated the transverse dimensions.

At first glance, our result (\ref{EMTresult}) contradicts that of 
Ref.~\cite{Kovchegov:2007pq}, where it was found that
the energy density should behave as a constant for $0<\tau\ll1$.
However, Ref.~\cite{Kovchegov:2007pq} based their analysis
on the assumption of boost-invariance, which is violated in our case,
and we have neglected transverse dynamics which could change 
the behavior of $\lim_{\tau\rightarrow 0}T_{\tau\tau}$. It may
thus be possible to reconcile our results with that of 
Ref.~\cite{Kovchegov:2007pq}. On the other hand, 
Ref.~\cite{Kovchegov:2007pq} forbid solutions of rising energy
density by invoking the positive energy condition 
$T_{\mu \nu} \omega^\mu \omega^\nu\ge 0$, where $\omega^\mu$
is a time-like vector (see also \cite{Janik:2005zt}). 
While this criterion is certainly
valid in classical gravity it is somewhat questionable why
it should apply to the EMT of the boundary quantum field theory.
After all, it is well-known that quantum fields cannot always and
everywhere satisfy all energy conditions \cite{Epstein}, and even the
weakest form of energy conditions, the averaged null energy condition, can
be violated for conformally coupled quantum fields in 3+1 dimensions in
any conformal quantum state \cite{Visser:1994jb}. It may be confusing
to find quantum effects in a classical gravity calculation: however,
it should be pointed out that these effects appear only at the boundary
of $AdS_5$, where the description should be dual to a strongly
coupled quantum field theory \cite{Maldacena:1997re}. Indeed,
the five-dimensional EMT in our calculation trivially fulfills
the positive energy condition, as it should for a classical gravity
calculation.


The breaking of boost-invariance of the EMT Eq.~\eqref{EMTresult} is a direct
consequence of the ansatz \eqref{EMTpre3} for the incoming shockwaves.
Namely, the relevant part of the precollision line element is
\beq
ds^2\sim z^2 \left[ \mu_1 \delta(x^-) dx^{-2}+
\mu_2 \delta(x^+) dx^{+2}\right],
\eeq
which -- in terms of $\tau,\eta$ coordinates --
has an explicit $\eta$ dependence. It is somewhat surprising,
though, that one can obtain a boost-invariant line element
when formally replacing $\delta(x)\rightarrow \partial_x \delta(x)$,
or similar structures\footnote{
Ref.~\cite{Kajantie:2008rx} proposed $\theta(x)/x^2$ to
obtain boost-invariance.}. In such a case, we could not
find a coordinate transformation similar to Eq.~(\ref{firstcotra})
in order to bring the line element into the Rosen form. 
It is conceivable that it would amount to replacing
$v \theta(v)$ and $u \theta(u)$ in Eq.~(\ref{superpo})
by their derivatives, in which case the line element
would be manifestly boost-invariant and the EMT
start from a finite value. However, it is not evident
how to interpret the physical meaning of 
a collision of two shock waves given by derivatives of delta functions.

\begin{figure}
\begin{center}
\includegraphics[width=\linewidth]{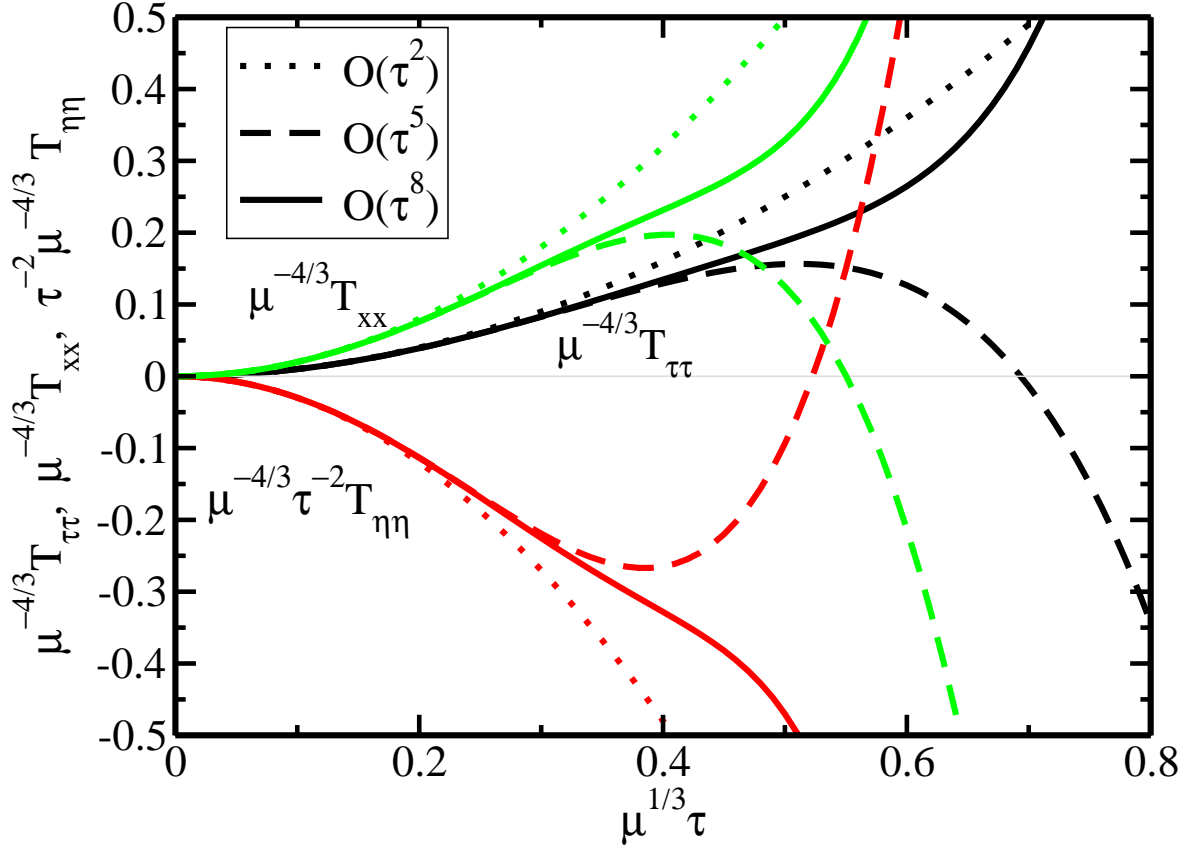}
\end{center}
\caption{The early time behavior of the individual components
of the EMT at $\eta=Y$.}
\label{fig1}
\end{figure}

In Fig.~\ref{fig1} we show 
a plot of the energy density and effective pressures in
the local rest-frame for early times, which suggest that our
small time expansion converges rapidly 
up to times of $\mu^{1/3} \tau\sim 0.4$.
At this time, the effective longitudinal pressure is still negative
and therefore the system is clearly not in equilibrium.
At times $\mu^{1/3} \tau\gtrsim 0.7$ the expansion in powers of $\tau$
seems to break down, possibly signaling the onset of a transition
to an equilibrated state, which from hydrodynamics is known
to behave as $T_{\tau \tau}\sim \tau^{-4/3}$.

It would be interesting to study in detail the horizon structure
of Eq.~(\ref{oursol}). Short of doing this, we may hope to learn
something about the position of the horizon by invoking cosmic
censorship, e.g.~requiring that all singularities of the 
metric (\ref{oursol}) are hidden behind horizons. To this end,
it is instructive to consider the minimum value of $\tilde z$ where
$H(\tilde \tau, \tilde \eta,\tilde z)=-1$ at early times.
Approximating $H$ by its first term, we expect a singularity
to appear at $\tilde z^4\le \frac{1}{2 \mu^2 \tau^2}$, which 
would have to be hidden by a horizon in order not to violate
cosmic censorship. Note that this implies the horizon is moving
towards the boundary at small times, in contrast to the situation
at late times studied in Ref.~\cite{Janik:2005zt}. Given that
in our case the energy density is initially rising and that for static
systems the temperature is inversely proportional to the 
distance of the black brane to the boundary, this is not too surprising.
As a caveat, it should be pointed out that 
$\tilde z^4\sim\frac{1}{\mu^2 \tau^2}$
is to be understood only as a very rough estimate of the 
singularity location since
for these values of $\tilde z$ essentially all terms of Eq.~(\ref{solfunc})
become of the same order, even at small times. 

Assuming that a singularity really does appear at 
$\tilde z^4\sim\frac{1}{\mu^2 \tau^2}$, how long would it take
until the EMT at the boundary $z=0$ ``knows'' about the
formation of the black hole (i.e.~the thermalization of the system)? 
We again estimate the time by assuming the information gets transported
over the distance $\tilde z$ with the speed of light, so 
$\tilde \tau\sim \tilde z$. Hence we obtain a crude estimate
of the thermalization time as a function of the transverse
energy density $\mu$,
\beq
\tau_{\rm therm}\sim \mu^{-1/3}\,,
\label{ttime}
\eeq
where the non-trivial dimensionless prefactor is ${\mathcal O}(1)$
in our simple estimate (from Fig.\ref{fig1}, one 
can extract $\tau_{\rm therm}>0.4 \mu^{-1/3}$). 
Taken at face value, this would imply extremely small
times $\tau_{\rm therm}\ll 1$ fm/c for modern colliders like RHIC or the LHC.
However, it seems that numerical simulations will be necessary to 
provide a detailed study of thermalization and 
extract the dimensionless prefactor in (\ref{ttime}) reliably.
Moreover, in a more realistic model than the one considered here,
we expect the dynamics in the transverse coordinates ${\bf x}_\perp$
to modify this thermalization time. We plan to study this 
in the near future \cite{inprep}.

\section*{Acknowledgments}

We thank R.~Janik, K.~Kajantie, A.~Karch, Y.~Kovchegov, R.~Peschanski, K.~Rajagopal, 
D.T.~Son and L.~Yaffe for fruitful discussions.
This work is supported in part by funds provided by the 
U.S. Department of Energy (D.O.E.) under the cooperative 
research agreement DEFG02-05ER41360 and by the Natural 
Sciences and Engineering Research Council of Canada. DG is 
supported by the project MC-OIF 021421 of the European 
Commission under the Sixth EU Framework Programme for 
Research and Technological Development (FP6). DG is grateful 
for the kind hospitality at the University of Washington where 
the current paper was initiated. The work of PR was 
supported by the US Department of Energy, grant 
number DE-FG02-00ER41132. PR would like to thank the 
Yukawa Institute for Theoretical Physics for 
the kind hospitality during the 
``Yukawa International Program for Quark-Hadron Sciences'',
where this work was being finished.

\appendix
\section{Distributions and Energy Conservation}
\label{mappendix}

In order to address distributional issues we make the following {\em global}
Ansatz for the line element in Rosen coordinates
%
\beq
ds^2=\frac{-2 du dv g_{1}(u,v,\tilde z)
+\left(u^2 dv^2+v^2 du^2\right) g_{2}(u,v,\tilde z)
+g_{3}(u,v,\tilde z) d{\bf x_\perp}^2 +
g_{4}(u,v,\tilde z){d\tilde{z}^2}}{\tilde z^2}\, .
\label{lel}
\eeq
For the precollision line element we can write
\bqa
g_{1}(u,v,\tilde z)=g_{3}(u,v,\tilde z)=f_1(u,\tilde z) f_1(v, \tilde z),\quad
g_{2}(u,v,\tilde z)=0,\quad
g_{4}(u,v,\tilde z)=f_2(u,\tilde z) f_2(v, \tilde z),\nonumber\\
f_1(u,\tilde z)=\left[1+2 \bar\mu \tilde{z}^2\ u\ \theta(u)\right]^{-2},
\quad 
f_2(u,\tilde z)=\left[1+6 \bar \mu \tilde{z}^2\ u\ \theta(u)\right]^{2}
\left[1+2\bar\mu \tilde{z}^2\ u\ \theta(u)\right]^{-2},
\nonumber
\eqa
where we have set $\mu_1=\mu_2=\bar{\mu}$ for simplicity.
For this line element, the Einstein equations (\ref{eq:Eeq}) do 
not contain any terms of the form $\delta(u),\delta(v)$ which 
would be singular at $u=0,v=0$. Nevertheless, one might be
worried that when making an ansatz for the line element for $u>0,v>0$
this would introduce ``spurious'' singularities. To test for this,
we choose the ansatz
\bqa
g_{1}(u,v,\tilde z)&=&f_1(u,\tilde z) f_1(v, \tilde z)
+{\mathcal O}(u v)
\nonumber\\
g_{2}(u,v,\tilde z)&=&\theta(u) \theta(v) \left(f_5(u,\tilde z)+f_5(v,\tilde z)\right)
+{\mathcal O}(u v)
\nonumber\\
g_{3}(u,v,\tilde z)&=&f_1(u,\tilde z) f_1(v, \tilde z)+
u v \theta(u) \theta(v) \left(f_6(u,\tilde z)+f_6(v,\tilde z)\right)
+{\mathcal O}(u^2 v^2)
\nonumber\\
g_{4}(u,v,\tilde z)&=&f_2(u,\tilde z) f_2(v, \tilde z)+
u v \theta(u) \theta(v) \left(f_7(u,\tilde z)+f_7(v,\tilde z)\right)
+{\mathcal O}(u^2 v^2),
\label{apdecomp}
\eqa
where $f_5(u,\tilde z),f_6(u,\tilde z),f_7(u,\tilde z)$ are required 
to be non-singular at $u=0$, and we neglected terms of higher order that 
result in regular terms in (\ref{eq:Eeq}).
Requiring that all terms of the form $\delta(u),\delta(v)$
cancel in the Einstein equations gives the condition
\bqa
\left(f_5(v,\tilde z)+f_5(0,\tilde z)\right) 24 \bar \mu^2 \tilde z^4  v^2 \theta(v)^2  
\frac{(1+6 \bar \mu \tilde z^2 v \theta(v) )}{(1+2 \bar \mu \tilde z^2 v \theta(v) )}
- f_7(v,\tilde z)-f_7(0,\tilde z)
\hspace*{2cm}\nonumber\\
-2 \left[f_6(v,\tilde z)+f_6(0,\tilde z)\right]
(1+6 \bar \mu \tilde z^2v \theta(v))^2=0,
\label{nosingcond}
\eqa
and likewise for $v\leftrightarrow u$.
If Eq.~(\ref{nosingcond}) is fulfilled, then the equations (\ref{eq:Eeq}) 
are regular at $u=0,v=0$ and can be conveniently solved in $\tau,\eta$-
coordinates.
By recasting our solution (\ref{oursol}) into the form (\ref{apdecomp}), we find
\bqa
f_5(v,\tilde z)+f_5(0,\tilde z)&=&\frac{1}{(1+2 \bar \mu \tilde z^2v \theta(v))^2} 
\lim_{u\rightarrow 0} \frac{L-K}{2 u v} \nonumber\\
f_6(v,\tilde z)+f_6(0,\tilde z)&=&\frac{1}{(1+2 \bar \mu \tilde z^2v \theta(v))^2}
\lim_{u\rightarrow 0} \left[\frac{8 z^4 \bar\mu^2}{(1+2 \bar \mu \tilde z^2v \theta(v))}
+\frac{H}{u v}\right]\nonumber\\
f_7(v,\tilde z)+f_7(0,\tilde z)&=&\frac{(1+6 \bar \mu \tilde z^2v \theta(v))^2}
{(1+2 \bar \mu \tilde z^2v \theta(v))^2}
\lim_{u\rightarrow 0} \left[\frac{8 z^4\bar\mu^2}{(1+2 \bar \mu \tilde z^2v \theta(v))}
-\frac{72 z^4\bar\mu^2}{(1+6 \bar \mu \tilde z^2v \theta(v))}
+\frac{M}{u v}\right].\qquad
\eqa
Inserting the expressions (\ref{solfunc}) with $\cosh[n(Y-\eta)]=\frac{2^{n-1}}{(\mu \tau)^n}
\left(u^n \bar\mu^n+v^n \bar\mu^n\right)$ and $\mu \tau= 2 \bar \mu \sqrt{u v}$
and reinstating appropriate $\theta$ functions for every appearance of $u,v$,
we obtain power series' in $v$ for the above $f_5,f_6,f_7$, respectively.
Using these, we have verified \emph{a posteriori}
that our solution obeys Eq.~(\ref{nosingcond}) order by order in $v$ (and by symmetry
also in $u$), and hence all $\delta$-functions in the Einstein equations cancel
to that order. Since (\ref{oursol}) is a solution to the Einstein equations
for arbitrarily small proper times, this implies that we know the metric
for $u\ll1,v\ll1$, including $u=0,v=0$.

As a consequence of that, we are able to calculate the EMT for small
(positive and negative) $x^\pm$ by repeating the holographic renormalization procedure of 
section \ref{sec:two} for the metric in the form (\ref{lel}) to Brinkmann
coordinates.
One finds 
\beq
T_{--}=-\frac{1}{2} \partial_-^2 a_0(x^+,x^-)+\frac{x^+}{x^-} \theta(x^+) \theta(x^-)
\lim_{z\rightarrow 0}\frac{L-K}{2 z^4}
\eeq
which using
\bqa
a_0(x^+,x^-)&=&-2 \bar \mu \left(x^+ \theta(x^+)+ x^- \theta(x^-)\right)
-\frac{5+c_1}{3} \bar \mu^2 4 x^{+2}x^{-2}\theta(x^+)\theta(x^-) 
+ {\mathcal O(x^{+3},x^{-3})}\nonumber\\
\lim_{z\rightarrow 0}\frac{L-K}{2 z^4}&=&-\frac{4}{3}(8+c_1) \bar\mu^2 x^+ x^- \theta(x^+)\theta(x^-)  + {\mathcal O(x^{+2},x^{-2})}
\eqa
becomes
\beq
T_{--}=\bar \mu \delta(x^-)
-4 x^{+2} \bar \mu^2 \theta(x^+)\theta(x^-) + {\mathcal O(x^{+3},x^{-})}.
\eeq
Note that in order for holographic renormalization to be consistent,
all terms proportional to $z^2$ in $L-K$ have to cancel.
We have explicitly verified this up to ${\mathcal O}(\tau^{10})$.
By symmetry we have 
\beq
T_{++}=\bar \mu \delta(x^+)
-4 x^{-2} \bar \mu^2 \theta(x^+)\theta(x^-) + {\mathcal O(x^{-3},x^{+})},
\eeq
and from (\ref{EMTresult}) we can glean
\beq
T_{+-}=\frac{1}{2}\left(T_{\tau \tau}-\tau^{-2}T_{\eta \eta}\right)
=8 x^+ x^- \bar \mu^2 \theta(x^+)\theta(x^-)+ {\mathcal O(x^{-2},x^{+2})}.
\eeq
This can be used to calculate the energy density
in the more familiar $t,z$ coordinates. The transformation is straightforward
and one finds
\bqa
T^{00}(t,z)&=&\frac{1}{2}\left(T^{++}+2 T^{+-}+T^{--}\right)\\
&\simeq&\frac{1}{2} \bar \mu \left(\delta(x^+)+\delta(x^-)\right)
-2 \bar \mu^2 \theta(x^+)\theta(x^-) \left(x^{+2}+x^{-2}-4 x^+ x^-\right) \nonumber\\
&\simeq&\frac{1}{\sqrt{2}} \bar \mu \left(\delta(t+z)+\delta(t-z)\right)
+2 \bar \mu^2 \theta(t^2-z^2)\left(t^{2}-3 z^2\right) \, .\nonumber
\eqa
Taking the result (to this order of calculation) at face value,
for fixed time $t$
the energy density deposited in the forward lightcone is positive 
at the collision point $z=0$ (corresponding to mid-rapidity $\eta=0$), 
while it becomes negative for some
$z^2\lesssim t^2$, especially close to the (positive) $\delta$-functions
at $z^2=t^2$.
We believe this is allowed for a quantum field theory, as discussed
in section \ref{sec:three}, but concede that 
the physical interpretation of this result demands further study.
Calculating the total energy at any given time $t$ gives
\beq
E(t)\equiv \int_{-\infty}^\infty dz T^{00}(t,z) = \sqrt{2}\bar \mu,
\eeq
up to the accuracy of the approximation. Therefore, 
the total system energy is conserved after the collision, 
which serves as yet another check on our result.


\end{document}